\documentclass[a4paper,11pt]{article}
\usepackage{jheppub} % for details on the use of the package, please see the JINST-author-manual
\usepackage{lineno}
\usepackage{float}
\usepackage{graphicx,xcolor}
\usepackage{enumitem}
\usepackage{braket}
\usepackage[colorlinks=true]{hyperref}
\usepackage{amsfonts,amssymb}
\usepackage{amsmath, mathtools}
\usepackage{caption,subcaption}

\usepackage[normalem]{ulem}

\usepackage{float}
\usepackage{graphicx,xcolor}
\usepackage{tikz}
\usetikzlibrary{arrows.meta}

\usetikzlibrary{positioning, calc}
\usepackage[edges]{forest}
\usepackage{enumitem}
\usepackage{braket}
\usepackage{comment}

\usepackage{amsthm}

\newcommand{\Tr}{\rm Tr}

\def \xr {\mathrm{x}}

\definecolor{amethyst}{rgb}{0.54, 0.17, 0.89}

\title{\boldmath {Detector-based measurement-induced state updates in AdS/CFT}}

\author[a,b,c]{Vijay Balasubramanian}
\emailAdd{vijay@physics.upenn.edu}
\affiliation[a]{David Rittenhouse Laboratory, University of Pennsylvania,  Philadelphia, PA 19104, USA}
\affiliation[b]{Theoretische Natuurkunde, Vrije Universiteit Brussel (VUB), B-1050 Brussels, Belgium}
\affiliation[c]{Rudolf Peierls Centre for Theoretical Physics, University of Oxford, Oxford OX1 3PU, UK}

\author[d,e,f]{Esko Keski-Vakkuri}
\emailAdd{esko.keski-vakkuri@helsinki.fi}
\affiliation[d]{Department of Physics, P.O.Box 64, FIN-00014 University of Helsinki, Finland}
\affiliation[e]{InstituteQ - the Finnish Quantum Institute, Finland}
\affiliation[f]{Helsinki Institute of Physics, P.O.Box 64, FIN-00014 University of Helsinki, Finland}

\author[f,g]{Nicola Pranzini}
\emailAdd{nicola.pranzini@helsinki.fi}
\affiliation[g]{QTF Centre of Excellence, Department of Physics, P.O.Box 64, FIN-00014 University of Helsinki, Finland}

\abstract{Conventional understandings of quantum theory hold that measurements change the state of an observed system following the L\"{u}ders update rule. Textbooks describe the application of this idea to non-relativistic systems, but extensions to relativistic and gravitating systems encounter subtleties.   One consistent approach is via detector-based measurements.  We study the effects of such measurements in a CFT with a holographic dual. We work out the boundary space-time regions associated to a L\"{u}ders update and how the outcome extends to modifications of the bulk gravity state.  We explore information-theoretic consequences of this picture, and relate the information extracted by a measurement to updates of the semiclassical parameters of the bulk state.}

\begin{document}
\maketitle
\flushbottom

\section{Introduction}
The idea that observers must be included within the quantum mechanical description of a system has been explored in various contexts, including the foundations of quantum theory~\cite{Zurek82, Zurek03, Zurek09}, quantum time~\cite{PageW83, GiovannettiEtAl15}, Quantum Reference Frames (QRF)~\cite{GiacominiEtAl19, Glowacki24, FewsterEtAl24},  algebraic approaches to quantum field theory in curved spacetime~\cite{ChandrasekaranEtAl23, Witten23, FewsterEtAl24}, observables in closed universes~\cite{ChandrasekaranEtAl23, Shaghoulian23}, and analyses of the gravitational path integral~\cite{AbdallaEtAl25, HarlowEtAl25, AkersEtAl25}. 
These approaches suggest that observers should not always be considered as external or factorised from the systems they observe. For example, the algebra of observables in de Sitter space is reduced from type ${\rm III}_1$ to the more tractable type ${\rm II}_1$ by including observers in the static patch as constraints on observables~\cite{ChandrasekaranEtAl23}. More generally, incorporating an observer---seen as a QRF---as an additional system in curved space Quantum Field Theory (QFT) leads to sufficient conditions, tied to properties of the QRF (and thus of the observer), for such a reduction in algebraic type \cite{FewsterEtAl24}. Similarly, the non-perturbative  Hilbert space required to describe observables accessible to gravitating observers exhibits non-trivial structure in closed universes, while the Hilbert space without observers seems to be one-dimensional and hence trivial~\cite{Balasubramanian:2023xyd,AbdallaEtAl25}. Broadly speaking, it is becoming increasingly clear that answers to questions posed independently of who is asking them may differ from those tied to the experience of observers and that the latter type of question may be more appropriate when gravity is involved. 

Here, we focus on the effects of measurements made by an observer. One motivation comes from recent work on measurements in holographic theories. There a particular problem is how measurements made by an observer in the bulk geometry can be given an interpretation in the fundamental description by the dual theory that lives in the boundary. The problem is particularly severe when the measurement is made behind a black hole horizon: is it even possible to find a description for the measurement in the fundamental theory, and if so, what are the observables, and what are the boundary state update rules given the measurement outcome? Ref.~\cite{Akers:2022qdl} proposed that the answer to these questions involves a non-isometric map from the bulk to the fundamental theory. This then implies that reconstruction of operators behind a horizon from the underlying theory must be state-dependent \cite{Antonini:2024yif}, and is likely computationally hard \cite{Akers:2022qdl,Balasubramanian:2022fiy}, as also expected from general considerations \cite{Brown:2019rox,Engelhardt:2021mue}. In fact, there are even more basic subtleties concerning the measurement of a state or operator in a quantum field theory: a too simple implementation of a local measurement-induced update rule leads to faster-than-light communication~\cite{Sorkin93}. This work focuses on measurements that are local, trigger a non-local update, and are realized by a specific choice of an instrument. Although measurements in holography have been explored before, previous approaches typically lacked one of these features. A non-exhaustive list includes the study of boundary non-local measurements in the presence of bulk black holes~\cite{ChowdhuryEtAl21}, how measurements change the boundary entanglement structure and thus the bulk's geometry~\cite{AntoniniEtAl22, AntoniniElAl23a, AntoniniEtAl23b}, and weak measurements at the boundary~\cite{SunJ23}.

We present a consistent model for measuring a CFT state at the boundary of AdS space, and investigate repercussions of a measurement for the dual bulk theory.  We identify  spacetime regions where the state update takes place, compute the information gained in the measurement, and show that this information is related to geometric quantities in the bulk. Most of the concepts we use have been discussed before, but usually in separate contexts. We assemble these concepts to discuss measurement in AdS/CFT in light of recent developments in the evolving theory of quantum measurements in relativistic theories.

\section{Detector-based measurements}

\subsection{Brief review of quantum measurements}

We begin with a short review of measurements as formulated in \cite{Zurek03}, and in \cite{Akers:2022qdl} in the holographic context. We have a quantum system $S$ of which we wish to measure a Hermitian observable $\hat{X}$. The measurement will reveal to us which of the eigenstates $\ket{n}_S$ of $\hat{X}$ the system will be in after the measurement. 
To achieve the measurement, we need an apparatus $A$ with a basis $\ket{x_n}_A$ pairing with the eigenstates $\ket{n}_S$ of $\hat{X}$. While the label $n$ need not be discrete, in practice, we can only measure a finite number of states due to restrictions on energy required by the measurement, measurement accuracy, and other technical limitations~\cite{PranziniV24}. We thus focus on measurements involving a finite range $n=0,\ldots ,N$ of states and measurement outcomes. 

The measurement begins with an interaction between the system and the apparatus, which can be described by some unitary operator $\hat{U}$ describing the evolution from the time the apparatus is brought in contact with the system to the time just before a reading of the apparatus is made. Before the interaction takes place, let us call the state of the apparatus $\ket{0}_A$ and the state of the system $\rho_S$. After the interaction $\hat{U}$, an observer reads the apparatus $A$ \footnote{Here we are omitting details of the observer being classical and having a pointer basis $\ket{n}_O$ matching the possible readings of the apparatus~\cite{Zurek82}. Recent discussions of this in the context of holography and quantum gravity can be found {\em e.g.} in \cite{Akers:2022qdl, HarlowEtAl25}.} and learns one of the $N+1$ measurement outcomes $x_n$. Therefore, one can introduce the measurement operators
\begin{equation}
    \hat{M}_n \equiv \bra{x_n}_A \hat{U} \ket{0}_A \label{measureM}
\end{equation}
pairing with the measurement outcomes, and satisfying the normalisation
\begin{equation}
    \sum^N_{n=0} \hat{M}^\dagger_n \hat{M}_n = \mathbb{I}_S~.
\end{equation}
As a result, the probability of measuring the outcome $x_n$ conditioned on the initial state $\rho_S$ of the system before the interaction and measurement is given by the Born rule
\begin{equation}
    p(x_n) = {\rm Tr}[\hat{M}^\dagger_n\rho_S \hat{M}_n]~.
\end{equation}
After knowing the outcome $x_n$, the state of the system is updated as
\begin{equation}
    \rho_S~\longrightarrow~\rho_n = \frac{\hat{M}_n\rho_S \hat{M}_n^\dagger}{p(x_n)}~; \label{Luders}
\end{equation}
this is known as the Lüders rule or, more commonly, as the post-measurement ``collapse'' of the state.  The Born rule and Lüders rule together are known as the measurement postulate. In the following, we consider the case where the system is initially in a pure state $\ket{\psi}_S$, so that \eqref{Luders} simplifies to 
\begin{equation}
    \ket{\psi}_S~\longrightarrow~\ket{n}_S = \frac{\hat{M}_n\ket{\psi}_S}{\sqrt{p(x_n)}}~. \label{Luders2}
\end{equation}

\subsection{Measuring a relativistic system}

Now we consider the system $S$ to be a quantum field. This presents a new problem: extending the measurement postulate to relativistic field theories, such as CFTs dual to AdS spacetimes, is itself non-trivial. In \cite{Sorkin93}, Sorkin demonstrated that naively extending non-relativistic measurement theory to QFT leads to faster-than-light communication. The result was obtained by adding the above measurement postulate to a minimal model of a relativistic theory, which includes the locality of observables, unitarity of time evolution, and microcausality of commutators. Then, requiring that the above state update applies  to all observers, measurements by observers with partially overlapping causal pasts/futures lead to faster-than-light communication~\cite{PapageorgiouF24}.  In response, various authors have tried to develop local measurement schemes to avoid Sorkin's problem. The various proposals differ in the assumed properties of the apparatus $A$, and the main alternatives are to take it to be another relativistic quantum field~\cite{FewsterV20} or a non-relativistic detector~\cite{Polo-GomezEtAl21}. In line with our simplifying assumption that there are a finite number of  measurement outcomes, we focus on the latter approach as a way to include measurements in QFT in a realistic manner. This approach is called  {\it detector-based measurement}; for example, we can utilise a non-relativistic two-level system,  the Unruh-DeWitt detector, as the apparatus~\cite{Polo-GomezEtAl21, PranziniKV23}.
This approach will also allow us to construct a specific unitary $\hat{U}$ implementing \eqref{measureM}.

While we expect that the Lüders rule \eqref{Luders} still applies in a relativistic theory, it is \textit{a priori} unclear which states or spacetime regions the update should affect. In fact, measurements are local operations, but the related state update is not. The only naturally covariant notions of a surface where the update could be applied are the past or the future light cones anchored at the measurement region. While an update at and inside the future light-cone may seem natural, consistency conditions, {\em e.g.}, for computing $n$-point functions, favour an update everywhere except in the causal past of the measurement region~\cite{PranziniKV23}. Importantly, faster-than-light communication remains impossible with this update rule, since single-shot measurement outcomes are inherently random and uncontrollable. Therefore, in the remainder of this work we assume that a measurement updates the state of a QFT  everywhere but in the causal past of the measurement region.  Some puzzles regarding this update rule remain under discussion, see, \textit{e.g.}, \cite{Polo-Gomez:2025nuy}.

\subsection{Unruh-DeWitt measurement apparatus}
\label{s.prepare_and_measure_UdW}
For our purposes it  suffices to take the apparatus to be the simplest two-level Unruh-DeWitt detector~\cite{Unruh76, DeWitt80}, with the free Hamiltonian 
\begin{equation}
    \hat{H}_\mathcal{D}=\Omega_g\ket{0}\bra{0}_A+\Omega_e\ket{1}\bra{1}_A~,
\end{equation}
where $\ket{0}_A$ and $\ket{1}_A$ are the orthogonal ground and excited states of the detector. An observer operating the detector moves along a trajectory $\xr(\tau)$ parameterised by  proper time $\tau$. The observer switches on and off the detector;  during this time window the detector interacts with a field operator $\hat{O}(x)$. A  common  model of this kind has the time-dependent interaction Hamiltonian 
\begin{equation}
    \hat{H}_{\text{int}}(\tau)=\lambda \chi(\tau)\hat{\mu}(\tau)\otimes\hat{O} (\xr(\tau))~,
    \label{e.H}
\end{equation}
where $\hat{\mu}(\tau)$ is the (interaction picture) operator
\begin{equation}
    \hat{\mu}(\tau) = \ket{1}\bra{0}_A e^{i(\Omega_e-\Omega_g) \tau} + \mathrm{h.c.}
\end{equation}
modeling energy level transitions in the detector, and $\chi (\tau)$ is the ``switching function'' modeling the switching on and off of the detector. The field operator is localised along the detector's trajectory.

We now have a detailed model for the system-apparatus interaction unitary operator in \eqref{measureM}, the time-evolution operator
\begin{equation} \label{e.time_evol_U}
    \hat{U}=\mathcal{T}\left\lbrace\exp\left[-i\int_{\tau_i}^{\tau_f} d\tau \hat{H}_{\text{int}}(\tau)\right]\right\rbrace~,
\end{equation}
where $\tau_i$ and $\tau_f$ are the proper times delimiting the support of $\chi(\tau)$. The measurement operators $\hat{M}_n$ are
\begin{equation}
    \hat{M}_0 \equiv \bra{0}_A \hat{U} \ket{0}_A \ ; \  \hat{M}_1 \equiv \bra{1}_A \hat{U} \ket{0}_A  \label{measureM2} \ .
\end{equation}
 Starting from the initial state where the detector is in the ground state and the QFT is in some generic pure state, $\ket{0}_A\otimes\ket{\psi}_S$, the state after the interaction reads
\begin{equation}
    \ket{\Psi_f}=c_0 \ket{0}_A\otimes\left(\hat{M}_0\ket{\psi}_S\right)+c_1 \ket{1}_A\otimes\left(\hat{M}_1\ket{\psi}_S\right)~.
    \label{e.final_state}
\end{equation}
The external drive injects energy to the system: as a result, the final state is a superposition of ground and excited states of the apparatus and the quantum field theory, and the two subsystems are entangled. If the quantum field theory in question is dual to bulk gravity as in AdS/CFT, then the state of the gravitating system  must also evolve into a superposition entangled with the (dual of the) detector. For example, if before the interaction the bulk was the AdS vacuum, this is no longer true after the interaction. Of course, this simple picture overlooks many details, including complications that emerge from energy non-conservation during single-shot measurements.

As an  example, choose the sharp switching function $\chi(\tau)=\delta(\tau-\tau_0)$ with $\tau_0\in[\tau_i,\tau_f]$.  We get 
\begin{equation}
   \hat{M}_{0}=\mathbb{I} ~~~ \textrm{and} ~~~ \hat{M}_{1}=-i\lambda e^{i(\Omega_e-\Omega_g)\tau_0}\hat{O}(x_M)
   \label{e.M}
\end{equation}
 at first order in perturbation theory in $\lambda$, where $x_M=\xr(\tau_0)$. 
Measuring the Unruh-De Witt detector by a projection  on the energy eigenbasis after the interaction induces a measurement on the field. The measurement-induced state collapse of the field should act as a selective update \eqref{Luders2} described by
\begin{equation}
    \ket{\psi}_S~\rightarrow\ket{\Phi_m}_S\propto~\begin{dcases}
        \ket{\psi}_S~~&{\rm if}~m=0\\
        %\hat{O}\ket{\Phi_0}~~&{\rm if}~m=1
        \hat{O}(x_M)\ket{\psi}_S~~&{\rm if}~m=1
    \end{dcases}~.
    \label{e.insertion_def}
\end{equation}
A natural guess is that this update should occur 
in the causal future of the measurement point $\mathcal{J}^+(x_M)$. This is almost correct. While it is clear that this update must at least happen in the causal future $\mathcal{J}^+(x_M)$ of the measurement,  non-relativistic quantum theory suggests that the update should be ``instantaneous''. In a relativistic setting, this can be understood by saying that the quantum measurement induces an update at all spacetime points that are simultaneous with (and in the future of) the measurement. As shown in \cite{PranziniKV23}, this is not just an implied consequence of a simple extension of non-relativistic quantum mechanics: stringent requirements from algebraic quantum field theory require the update to take place everywhere within the causal complement of the measurement’s past, {\em i.e.,} the complement of $\mathcal{J}^-(x_M)$. Specifically, consistent assignment of $n$-point functions requires the updated state to be valid at all spacetime points not in the causal past of the measurement.

So far, we have focused on single-shot measurements, which involve observing the outcome of a single run of the quantum measurement process. Experimentally relevant are also  many-shot measurements, repeating the same measurement multiple times and collecting all outcomes to build statistical distributions. For the present QFT setting, this becomes somewhat hypothetical, as it would mean taking many identical and independent copies of the spacetime together with the field configuration, and collecting statistics from such an ensemble. Disregarding the issue of practical implementation, many-shot measurements can be interpreted as non-selective measurements.  In a non-selective measurement, the updated state is the classical statistical average over the entire ensemble of single-shot post-measurement states. While single-shot measurements may misleadingly suggest the possibility of faster-than-light signalling, actual communication always involves many measurements and must be treated within the non-selective framework, thus preserving causality~\cite{AsherT04}. In the following, we will mostly consider single-shot experiments, but many-shot scenarios are also discussed when relevant.

We also focused on point-like detectors, but general models may also include a smearing function in space, thus giving the detector a spacelike extent~\cite{Schlicht04}, which we will take to be compact. Such models then describe an observer with a time-like world-line and carrying a detector to indirectly measure an operator $\hat{O}$ localised in the time-like tube surrounding its trajectory, connecting with the discussion in \cite{Witten23}. A smeared interaction could be regarded as simply a regularisation method, but can also be given a physical interpretation.   One possibility is to quantize the COM degree of freedom itself~\cite{FooEtAl20}.   An alternative is to consider the detector to have a classical center of mass (COM) which has a definite trajectory, but also additional internal quantum degrees of freedom, causing fluctuations about the detector's classical trajectory~\cite{LoppMM21}.  

Smearing the detector in either way smears its past light-cone. Consider a setting with additional internal degrees of freedom where different readings of the detector not only produce a measurement outcome $m$, but also produce a definite value for the measurement point $x_M$ and thus a definite past light-cone $\mathcal{J}^-(x_M)$. Since the outcome $x_M$ is random, different runs of the measurement will not always yield the same past light cone $\mathcal{J}^-(x_M)$.  We can then summarise the outcomes of the many runs of the experiment by a non-selective density operator~\cite{Polo-GomezEtAl21}.  For smeared detectors, these states will differ outside different light cones. Furthermore, a simple calculation shows that expectation values of local observables spacelike-separated from the measurement point(s) computed with respect to the non-selective state match with the expectation values computed with respect to the state before the measurement (\textit{i.e.}, within the intersection of all possible past light-cones)!  This is a field-theoretic version of the no-signalling theorem~\cite{AsherT04}.  Thus, in the case of many-shot measurements ({\em i.e.,} repeated experiments), there is no need for a sharp cut between the past of the measurement and the rest of the spacetime. This way, the state smoothly becomes the non-selective one when moving from the past of the measurement to its spacetime complement. In summary, two seemingly unrelated facts---the no-signalling theorem and the smearing of detectors---conspire to remove the need for a sharp cut between the past of the measurement and the rest of the spacetime for many-shot experiments.

We could  consider measurements that do not entirely localise the detector.  In that case, the light cone remains smeared even after an outcome is produced.  A concrete example of this is the case where the field interacts with the electron in the 1s orbital of a hydrogen atom (now taken to be the detector). Assuming that the interaction drives the detector and the field in a state where the 1s and 2s orbitals are respectively entangled with some field states, measuring the internal energy is not enough to collapse the detector into a classical trajectory. Instead, the post-measurement state will be a coherent superposition of positions for the detector, field states for the field, and causal pasts for the measurement. We will return to this scenario in the conclusion.

\section{Measuring QFT states}

We now consider the Unruh-DeWitt detector moving in a two-dimensional planar spacetime with metric 
\begin{equation}
    ds^2=-dt^2+dx^2
\end{equation}
and coupled to a quantum field theory (QFT).  Generalisations are straightforward, and, in particular, the QFT may or may not be a conformal field theory (CFT) with a holographic dual. Our primary goal here is to develop a framework for describing state updates using standard QFT language, rather than the algebraic formalism employed in \cite{PranziniKV23}.

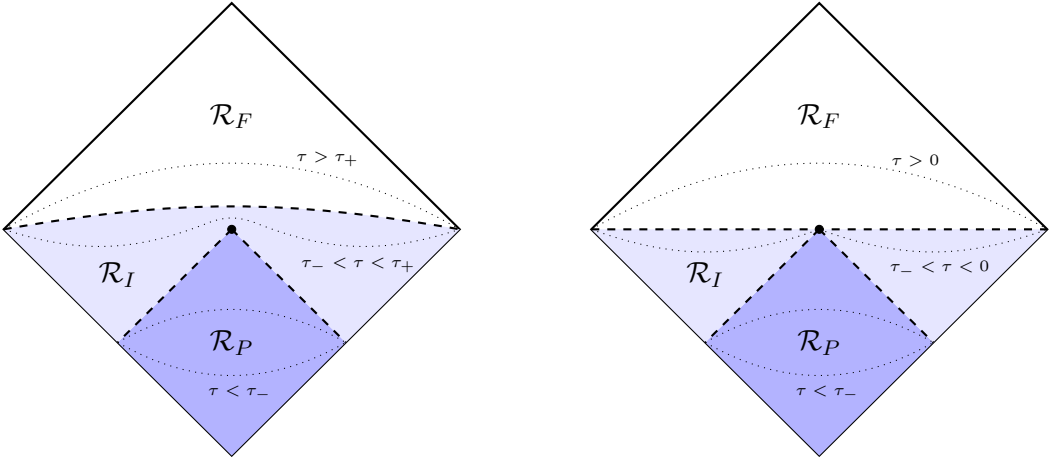
\begin{figure*}[t!]
    \centering
    \begin{subfigure}[b]{0.5\textwidth}
        \centering
        \begin{tikzpicture}[scale=3]

  % Coordinates
  \coordinate (iMinus) at (0,-1);      % i^-
  \coordinate (iPlus) at (0,1);        % i^+
  \coordinate (iZeroLeft) at (-1,0);   % i^0 (left)
  \coordinate (iZeroRight) at (1,0);   % i^0 (right)
  \coordinate (origin) at (0,0);       % origin

  % Penrose diagram boundary
  \draw[thick] (iMinus) -- (iZeroLeft) -- (iPlus) -- (iZeroRight) -- cycle;

  % Region II: I^-(0), shaded
  \fill[blue!30] (iMinus) -- (-0.5,-0.5) -- (0,0) -- (0.5,-0.5) -- cycle;
  \fill[blue!10]
  (origin) -- (-0.5,-0.5) -- (iZeroLeft)
  to[bend left=10] (iZeroRight) -- (0.5,-0.5) -- cycle;

  % Axes (optional)
  \draw[thick, dashed] (iZeroLeft) to[bend left=10] (iZeroRight); % t = 0
  %\draw[dashed] (0,-1) -- (0,1);             % x = 0

  % Lightcones from origin
  \draw[thick, dashed] (0,0) -- (-0.5,-0.5);
  \draw[thick, dashed] (0,0) -- (0.5,-0.5);

  % slicing
  \draw[dotted] (-0.5,-0.5) to[bend left=30] (0.5,-0.5);
  \draw[dotted] (-0.5,-0.5) to[bend right=30] node[below, xshift=3pt] {\tiny $\tau < \tau_-$} (0.5,-0.5);

  \draw[dotted] 
  (iZeroLeft) to[bend right=20] (-0.1,0.03)
              to[bend left=20] (0.1,0.03)
              to[bend right=20] node[pos=0.7, below, xshift=-14pt, yshift=-1pt] {\tiny $\tau_- < \tau < \tau_+$} (iZeroRight);

  \draw[dotted] (iZeroLeft) 
  to[bend left=30] node[pos=0.7, above, xshift=2pt, yshift=-1pt] {\tiny $\tau > \tau_+$} 
  (iZeroRight);

  % Points
  %\node at (0,1.08) {$i^+$};
  %\node at (0,-1.1) {$i^-$};
  %\node at (-1.1,0) {$i^0$};
  %\node at (1.1,0) {$i^0$};
  \filldraw (origin) circle (0.5pt) node[below right] {};

  % Labels
  \node at (0,-0.5) {$\mathcal{R}_P$};
  \node at (-0.5,-0.2) {$\mathcal{R}_I$};
  %\node at (0.5,-0.2) {$\mathcal{R}_I$};
  \node at (0,0.5) {$\mathcal{R}_F$};

  %\foreach \y in {1, 2, 4} {
  %  \draw[domain=0:1, samples=100, smooth, variable=\t, blue, thick] 
  %plot ({-1+2*\t}, {\y*(cosh(0.5)-cosh(\t-0.5))});
  % }

\end{tikzpicture}
    \caption{Conformal diagram showing the regions and slices discussed in the text. The differently coloured regions are $\mathcal{R}_P=\mathcal{J}^-(x_M)$, and $\mathcal{R}_I\cup \mathcal{R}_F=\mathcal{M}\setminus \mathcal{R}_P$. Slices are represented as dotted lines, with $\tau>\tau_+$ being the usual Minkowski ones. Slices with $\tau_-<\tau<\tau_+$ can be everywhere smooth. }
    \label{f.regions}
    \end{subfigure}%
    ~ 
    \begin{subfigure}[b]{0.5\textwidth}
        \centering
        \begin{tikzpicture}[scale=3]

  % Coordinates
  \coordinate (iMinus) at (0,-1);      % i^-
  \coordinate (iPlus) at (0,1);        % i^+
  \coordinate (iZeroLeft) at (-1,0);   % i^0 (left)
  \coordinate (iZeroRight) at (1,0);   % i^0 (right)
  \coordinate (origin) at (0,0);       % origin

  % Penrose diagram boundary
  \draw[thick] (iMinus) -- (iZeroLeft) -- (iPlus) -- (iZeroRight) -- cycle;

  % Region II: I^-(0), shaded
  \fill[blue!30] (iMinus) -- (-0.5,-0.5) -- (0,0) -- (0.5,-0.5) -- cycle;
  \fill[blue!10] (origin) -- (-0.5,-0.5) -- (iZeroLeft) -- cycle;
  \fill[blue!10] (origin) -- (0.5,-0.5) -- (iZeroRight) -- cycle;

  % Axes (optional)
  \draw[thick, dashed] (iZeroLeft) -- (iZeroRight); % t = 0
  %\draw[dashed] (0,-1) -- (0,1);             % x = 0

  % Lightcones from origin
  \draw[thick, dashed] (0,0) -- (-0.5,-0.5);
  \draw[thick, dashed] (0,0) -- (0.5,-0.5);

  % Points
  %\node at (0,1.08) {$i^+$};
  %\node at (0,-1.1) {$i^-$};
  %\node at (-1.1,0) {$i^0$};
  %\node at (1.1,0) {$i^0$};
  \filldraw (origin) circle (0.5pt) node[below right] {};

  % Labels
  \node at (0,-0.5) {$\mathcal{R}_P$};
  \node at (-0.5,-0.2) {$\mathcal{R}_I$};
  %\node at (0.5,-0.2) {$\mathcal{R}_I$};
  \node at (0,0.5) {$\mathcal{R}_F$};

    % slicing
  \draw[dotted] (-0.5,-0.5) to[bend left=30] (0.5,-0.5);
  \draw[dotted] (-0.5,-0.5) to[bend right=30] node[below, xshift=3pt] {\tiny $\tau < \tau_-$} (0.5,-0.5);

  \draw[dotted] (iZeroLeft) to[bend right=20] (0,0);
  \draw[dotted] (0,0) to[bend right=20] node[pos=0.7, below, xshift=-14pt, yshift=-1pt] {\tiny $\tau_- < \tau < 0$}  (iZeroRight);

  \draw[dotted] (iZeroLeft) 
  to[bend left=30] node[pos=0.7, above, xshift=2pt, yshift=-1pt] {\tiny $\tau > 0$} 
  (iZeroRight);

  %\foreach \y in {1, 2, 4} {
  %  \draw[domain=0:1, samples=100, smooth, variable=\t, blue, thick] 
  %plot ({-1+2*\t}, {\y*(cosh(0.5)-cosh(\t-0.5))});
  % }

\end{tikzpicture}
    \caption{Simplified regions and slicing. We take $\tau_+\to 0$. The light-like interface $\gamma$ between $\mathcal{R}_P$ and $\mathcal{R}_I$ is called $\Sigma(\tau_-)$ and used to make $\Sigma(\tau)$ have domain $\mathbb{R}$. The slices in $\mathcal{R}_P$ are Milne slices, and those in $\mathcal{R}_I$ are the usual Rindler ones.\\
    ~\\
    ~}
    \label{f.regions with singular slices}
    \end{subfigure}
    \caption{Measurement-defined regions and compatible slicings contained therein.}
\end{figure*}

Suppose that unitary time-evolution of the quantum field and detector in Minkowski space brings an initial state (say, the Minkowski vacuum) into the entangled superposition \eqref{e.final_state}. Observers with access to the detector read it and learn the measurement outcome, say  $m=1$. Then the field must be updated to be in the excited state.

Now consider the interior of the past lightcone of the measurement point $\mathcal{J}^-(x_M)$ (which we call $\mathcal{R}_P$) and its complement.  We know that in the interaction picture, the updated state in the complement of $\mathcal{R}_P$ must be the excited state, while in  $\mathcal{R}_P$, it must be the initial field state $\ket{\psi}_S$, or rather the resulting reduced density matrix $\rho_S$ appropriate to $\mathcal{R}_P$.  A more precise statement is that the field is described by an algebraic local state $\omega$. Local algebraic states are linear maps from the algebra of local operators to the set of non-negative real numbers. Here the relevant algebra is $\mathcal{A}(\mathcal{R}_P)$ of local operators in the region $\mathcal{R}_P$ and the local algebraic states are the maps $\omega:\mathcal{A}(\mathcal{R}_P) \rightarrow \mathbb{R}^+$ such that for every local operator $A$ in that region its image under $\omega$ agrees with the vacuum expectation value, $\omega (A)= \Tr_{\mathcal{R}_P} (\rho_S A) = \bra{\psi}_SA\ket{\psi}_S$. The same is true for the expectation values of operators in the complementary region, with respect to the local algebraic states $\Tilde{\omega}$ prepared by the measurement therein, $\Tilde{\omega}(B)=\bra{\psi}_S\hat{O}^\dagger(x_M)B\hat{O}(x_M)\ket{\psi}_S$ for $m=1$.

In order to get a more complete understanding of the states, we move back to the Schrödinger picture. To restore the free time evolution in the updated states, we introduce new coordinates and constant time slices, so that on every slice we have a well-defined state, and a complete description of time evolution for every region.

The standard approach to field quantisation is to foliate the spacetime with a set of spacelike slices. However, this causes problems when the field state is being measured, for a generic slicing may cut through both $\mathcal{J}^-(x_M)$ and its complement, where the field state has potentially different assignments. Hence, one must construct an appropriate family of spacelike slices that remain within one causal region defined by the measurement at a time, so that on each constant time slice, the time-evolved state is unique.

Considering point-like measurements---hereafter localised at the origin $x_M=(0,0)$---and flat Minkowski spacetime, let us consider the regions $\mathcal{R}_P=\mathcal{J}^-(x_M)$ and its complement, which we further split into:
\begin{equation}
    \begin{split}
        \mathcal{R}_F&=\{(t,x)~|~t\geq t_+\}\\
        \mathcal{R}_I&=\{(t,x)~|~-|x|<t< t_+\}
    \end{split}
\end{equation}
with $t_+$ some arbitrary positive time. Also, we call $\gamma$ the light-like interface between $\mathcal{R}_P$ and $\mathcal{R}_I$. These regions are shown with different colours in Fig.~\ref{f.regions}. In the Heisenberg picture, the field has one value in $\mathcal{R}_P$ and another in $\mathcal{R}_I\cup\mathcal{R}_F$~\cite{PranziniKV23}. Next, we pick two positive numbers $\tau_+$ and $\tau_-$ and construct a family of spacelike slices $\Sigma(\tau)$ parametrised by $\tau\in\mathbb{R}\setminus\{\tau_-\}$, spanning the regions one by one. We impose $\tau_+=t_+$, and that $\Sigma(\tau\geq\tau_+)$ is the usual Minkowski coordinate time slicing and $\Sigma(\tau_-)$ is a light like surface matching $\gamma$. A representative slicing of this form is shown in Fig.~\ref{f.regions}. A similar  construction for a cylindrical spacetime is outlined in App.~\ref{a.slicing_construction}. 

For simplicity, we set $\tau_+ \to 0$  (Fig.~\ref{f.regions with singular slices}). In this case, the slices in $\mathcal{R}_I$ ($\tau_-<\tau<0$) may be chosen as those conventionally used to span the lower half left and right Rindler wedges. We can also use Rindler-Milne coordinates to define $\Sigma(\tau)$ in $\mathcal{R}_P$: slices with $\tau<\tau_-$ are defined as constant conformal time slices of a contracting Milne universe with coordinates $(\eta,\xi)$, related to $(t,x)$ via
\begin{equation}
    \begin{dcases}
        t=-e^{\eta}\cosh\xi\\
        x=-e^{\eta}\sinh\xi
    \end{dcases}
\end{equation}
with $\eta,\xi\in\mathbb{R}$~\cite{OlsonR11}.  In fact, slices with $\tau<\tau_-$ can be related to those with fixed conformal time $\eta$ via $\tau=\tau_--e^{\eta}$~\footnote{While having $\tau$ a decreasing function of $\eta$ may seem counter-intuitive, this is the usual choice made in the literature about QFT in the Milne universe and in past/future Minkowski wedges.}. With this convention, the future-directed Killing vector field $K_1$ corresponding to constant $\tau$ slices comes with a minus sign $K_1=-\partial_\eta$. The Milne metric also has another future-directed Killing vector field $K_2$, corresponding to $K_2=\partial_t$ in Minkowski coordinates.

Now consider the non-trivial case where we measured $m=1$ so that
we need to implement the update 
\begin{equation}
    \ket{\psi}_S\rightarrow\hat{O}(x_M)\ket{\psi}_S 
    \label{dummy}
\end{equation}
in $\mathcal{R}_{I}\cup\mathcal{R}_F$. In the Heisenberg picture, the updated state $\hat{O}(x_M)\ket{\psi}_S$ is fixed. Now we move to the Schrödinger picture, and restore the time evolution. Consider first the region $\mathcal{R}_F\cup\mathcal{R}_{I}$. We can take the state to be $\hat{O}\ket{\psi}_S$ at a slice $\Sigma (\tau)$ infinitesimally close to the measurement point and to the boundary of $\mathcal{R}_P$, and then follow the time-evolution into the future along first the Rindler and then the Minkowski slicing $\Sigma (\tau>0)$.

In region $\mathcal{R}_P$, the Schrödinger picture description is more complicated: the time-evolution now applies to the reduced state $\rho_S$ with the Hamiltonian written in Milne coordinates, corresponding to evolution generated by $K_1$. In general, both the reduced state and its time-evolution are difficult to work out explicitly. The simplest case is when the initial state is the Minkowski vacuum, $\ket{\psi}_S=\ket{0}_M$, in which case the reduced state in $\mathcal{R}_P$ in the chosen Milne foliation is a thermal state~\cite{OlsonR11}. Furthermore, the thermal state is also invariant under the time evolution generated by $K_1$ (see App.~\ref{a.CFT_details} for details). For simplicity, we will illustrate the ideas with this initial state.

So far, we have been discussing what is known as a prepare-and-measure setting.
Another interesting case is the task of distinguishing between two states. Suppose we know that the quantum field is in either one of two alternative states $\ket{\psi_1},\ket{\psi_2}$ but we do not know which one. This requires designing a measurement that can distinguish the states. In fact, local measurements of the sort described above generally cannot fully discriminate between states of a QFT.
This is due to the entanglement properties of QFT~\cite{SummersW87, Witten18}, for which most global pure states are mixed when seen from the perspective of a local observer. Thus, to distinguish between states, a local observer must compare their relative density operators by trace distance: two states are perfectly locally distinguishable only if the trace is null~\cite{Helstrom76}. But QFT reduced states do not have vanishing overlap unless specific physical conditions are realised (\textit{e.g.}, a wave packet is sent towards the observer such that it hits precisely during the measurement coupling). For this reason, state distinguishability for local observers is not easily addressed in general, and we therefore refrain from further comment. 

We need one more observation before discussing the bulk interpretation of the state update in AdS/CFT. Note that the update \eqref{dummy} %for $m=1$ 
superficially resembles a local operator quench \cite{NozakiEtAl13, NozakiEtAl14}, where a QFT state is suddenly acted on by a local operator. This resemblance is partially an artifact of using the simple sharp switching function $\chi(\tau)=\delta (\tau -\tau_0)$. To emphasize the different interpretation here, imagine that the detector is switched on adiabatically with $\chi (\tau) = \Theta (\tau_0-\tau) e^{(\tau-\tau_0)/\tau_{\textrm{large}}}$ so that the initial separable state $\ket{0}_A \otimes \ket{\psi}_S$ time-evolves slowly to the coherent entangled superposition \eqref{e.final_state}. It is clear that this prepared state is different from the state $\ket{\psi}_S$ before a local operator quench. Thus, rather than acting as a local operator quench, the measurement removes the (quantum) uncertainty about the state of the system, and the state is updated, conditioned by the measurement outcome.
Strictly speaking, it would also not be correct to call this an ``operation insertion'' at $x_M$, since the operator was already present in the prepared state before the measurement. Once the state has been updated, its time evolution in the Schrödinger picture depends on the choice of the operator that has been interacting with the detector. In Section \ref{sec:AdS-CFT} we consider the case of a heavy primary operator in CFT, where the subsequent time-evolution corresponds to a massive particle falling in the bulk.

\section{Measurements in AdS/CFT}\label{sec:AdS-CFT}

Having discussed detector-based measurements in QFTs, we now examine how these translate to the dual AdS bulk spacetime via the AdS/CFT correspondence. Within its domain of validity, this correspondence maps bulk field states in AdS to boundary CFT states, and vice versa.  In the appropriate limit, superpositions of CFT states can be read via the correspondence as superpositions of bulk field configurations and geometries. Thus, it is conceivable that measurement-induced updates correspond to sharp, non-local transitions between semiclassical bulk states or spacetime configurations. This section explores this possibility.

We begin by recalling the setup we use. We consider three systems in the boundary theory: (i) the system $S$ to be measured, which is some (holographic) CFT with Hilbert space $\mathcal{H}_S$; (ii) the measuring apparatus $A$, which is the UDW detector, \textit{i.e.}, an additional system external to the CFT with two-dimensional Hilbert space $\mathcal{H}_{A}$; and (iii) a classical observer who operates the detector. We invoke the Heisenberg cut to enforce the classical nature of the observer, thus treating them as classical and external to the quantum system. The detector's readout is hence modelled as a projective measurement in its energy eigenbasis, $\ket{0}_A,\ket{1}_A$, with outcomes conditioned on the classical information obtained by the observer. Finally, the total Hilbert space is the tensor product $\mathcal{H}_A\otimes\mathcal{H}_S$, where we left out the (pointer) states of the observer thanks to their required classicality.  Overall, this picture is in line with the recent literature\footnote{A further refinement to the discussion would be to consider the detector as a quantum reference frame and apply a relativisation map to the observables of the CFT, but that is beyond the scope of this work.} about measuring observers in (quantum) gravity~\cite{HarlowEtAl25}.

%A further refinement would be to consider the detector as a quantum reference frame, and apply a projection, {\em i.e.,} a relativisation map to $\mathcal{H}_A\otimes\mathcal{H}_S$, leaving a gauge invariant subspace of states. Since this step is not relevant for this manuscript, we leave it for future work. \nic{[Check by Esko, and add Eef.s; I dont' feel competent enough to comment on this yet.]}

The measuring protocol is as follows. The initial state is $\ket{0}_A\otimes \ket{\psi}_S$ where, to avoid complications arising from having event horizons in the bulk, we assume the initial state $\ket{\psi}_S$ of the CFT to be created from the Minkowski vacuum by a CFT operator of sub-exponential complexity.
The time-dependent Hamiltonian \eqref{e.H} couples the detector to a CFT operator $\hat{O}(x)$, which we also assume to be of sub-exponential complexity. After the interaction, the state of the detector and the CFT has evolved into an entangled superposition \eqref{e.final_state}, and the observer reads the detector. This is modelled as a conditional application of a projection $\ket{m}\bra{m}_A\otimes\mathbb{I}_S$, with $m=0,1$, 
onto \eqref{e.final_state}, which in turn results in a Lüders update of the CFT state, dual to an update of the bulk state. The above constraints on the initial state and coupling complexity mean we start and end without event horizons in the bulk, thus enabling the use of causal wedge reconstruction techniques (see later).

While the above focuses on local measurements in the boundary theory, the same applies when analysing a bulk apparatus localised near the boundary. With this setting in mind, we address the following questions:
\begin{enumerate}
    \item \textit{Where does the dual non-local state update occur in the bulk? More generally, how does a measurement-induced state update on the boundary manifest in the bulk?}
    \item \textit{What quantities relate the boundary measurement operation to the resulting bulk state? How can one quantify the change in the bulk state by quantities accessible at the boundary?}
\end{enumerate}

\begin{figure}
    \centering
    \begin{tikzpicture}[scale=2, x={(0.866cm,-0.5cm)}, y={(0cm,1cm)}, z={(0.5cm,0.4cm)}]

  % x–t plane (z = 0)
  \fill[blue!10, opacity=0.5] 
    (-1,-1,0) -- (2,-1,0) -- (2,2,0) -- (-1,2,0) -- cycle;
  \node at (1.8,1.8,0) {$\partial\mathcal{M}$};

  \node at (-1,-1,-2) {$\mathcal{M}$};

  \node at (-0.5,1.2,0) {$\mathcal{D}$};
  \node at (-0.5,1.2,-1.2) {$\mathcal{W}_\mathcal{D}$};

  % Axes
  \draw[->] (-1,0,0) -- (2,0,0) node[anchor=north east]{$x$};
  \draw[->] (0,-1,0) -- (0,2,0) node[anchor=north west]{$t$};
  %\draw[->] (0,0,0) -- (0,0,-3) node[anchor=north east]{bulk direction};

  \filldraw[black] (0,0,0) circle (1pt) 
    node[anchor=south west] {$x_M$};

  % Point off the plane (negative y direction)
  \filldraw[red] (-0.5,1,-0.3) circle (1pt) 
    node[anchor=west] {$p$};

% 1. Draw cone side lines (rays)
\foreach \angle in {180,190,...,360} {
    \ifnum\angle=180
        \draw[black, dashed, opacity=1] 
            (0,0,0) -- ({1*cos(\angle)}, {-1}, {1*sin(\angle)});
    \else
        \ifnum\angle=360
            \draw[black, dashed, opacity=1] 
                (0,0,0) -- ({1*cos(\angle)}, {-1}, {1*sin(\angle)});
        \else
            \draw[gray, thin, opacity=0.5] 
                (0,0,0) -- ({1*cos(\angle)}, {-1}, {1*sin(\angle)});
        \fi
    \fi
}

% 2. Fill cone surface (fan shape from 180° to 360°)
\fill[gray, opacity=0.1] 
  (0,0,0)
  \foreach \angle in {360,350,...,180} {
    -- ({1*cos(\angle)}, {-1}, {1*sin(\angle)})
  } -- cycle;

\coordinate (tipZ) at (-0.5,1.5,0);
\coordinate (tipZ2) at (-0.5,0.5,0);

% Second cone side lines (rays), same half-cone angles 180° to 360°
\foreach \angle in {180,190,...,360} {
    \ifnum\angle=180
        \draw[black, dashed, opacity=1] 
            (tipZ) -- ($ (tipZ) + ( {0.5*cos(\angle)}, {-0.5}, {0.5*sin(\angle)} ) $);
    \else
        \ifnum\angle=360
            \draw[black, dashed, opacity=1] 
                (tipZ) -- ($ (tipZ) + ( {0.5*cos(\angle)}, {-0.5}, {0.5*sin(\angle)} ) $);
        \else
            \draw[gray, thin, opacity=0.5] 
                (tipZ) -- ($ (tipZ) + ( {0.5*cos(\angle)}, {-0.5}, {0.5*sin(\angle)} ) $);
        \fi
    \fi
}

% Fill second cone surface
\fill[gray, opacity=0.1] 
  (tipZ)
  \foreach \angle in {360,350,...,180} {
    -- ($ (tipZ) + ( {0.5*cos(\angle)}, {-0.5}, {0.5*sin(\angle)} ) $)
  } -- cycle;

  % Third cone side lines (rays), same half-cone angles 180° to 360°
\foreach \angle in {180,190,...,360} {
    \ifnum\angle=180
        \draw[black, dashed, opacity=1] 
            (tipZ2) -- ($ (tipZ) + ( {0.5*cos(\angle)}, {-0.5}, {0.5*sin(\angle)} ) $);
    \else
        \ifnum\angle=360
            \draw[black, dashed, opacity=1] 
                (tipZ2) -- ($ (tipZ) + ( {0.5*cos(\angle)}, {-0.5}, {0.5*sin(\angle)} ) $);
        \else
            \draw[gray, thin, opacity=0.5] 
                (tipZ2) -- ($ (tipZ) + ( {0.5*cos(\angle)}, {-0.5}, {0.5*sin(\angle)} ) $);
        \fi
    \fi
}

% Fill third cone surface
\fill[gray, opacity=0.1] 
  (tipZ2)
  \foreach \angle in {360,350,...,180} {
    -- ($ (tipZ) + ( {0.5*cos(\angle)}, {-0.5}, {0.5*sin(\angle)} ) $)
  } -- cycle;

  \draw[thick, blue, opacity=0.7]
  plot[variable=\theta, domain=180:360, samples=50] 
    (
      { -0.5 + 0.5*cos(\theta) }, % x-coordinate
      1.0,                       % y-coordinate (fixed)
      { 0 + 0.5*sin(\theta) }    % z-coordinate
    );

\end{tikzpicture}
    \caption{An example of the casual wedge $\mathcal{W}_\mathcal{D}$ in the AdS spacetime $\mathcal{M}$ and the related projection $\mathcal{D}$ on the AdS boundary  $\partial\mathcal{M}$. We can reconstruct the field at any bulk point $p\in\mathcal{W}_\mathcal{D}$ such that the intersection between $\mathcal{W}_\mathcal{D}$ and the bulk past lightcone of $x_M$ is null. %As it is clear, it is always possible to select $\mathcal{W}_\mathcal{D}$ as such.
    }
    \label{f.causal_wedge}
\end{figure}
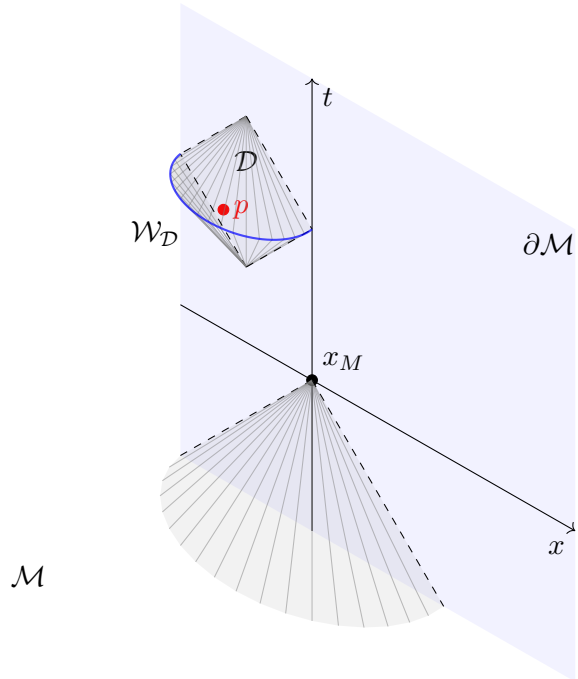

We begin with the first point. Once the regions defined by the boundary measurement have been identified, the update can be extended to an update on the bulk field's state dual to the CFT state. In the simple case of large $N$ and weakly coupled bulk limit, this can be done via the HKLL causal wedge reconstruction~\cite{HKLL06_2, HKLL06, HeemskerkEtAl12}. The only caveat is that one must choose the causal wedge $\mathcal{W}_D$ carefully. The causal diamond $D$ which supports the wedge must not intersect the lightlike surface $\gamma$ which emanates from the measurement point, and the bulk wedge $\mathcal{W}_D$ must not intersect the light-like surface anchoring at $\gamma$. Otherwise, the state inside $D$ is not uniquely defined, and the bulk update becomes ambiguous inside $\mathcal{W}_D$. Thankfully, since the cut between the measurement-defined regions is light-like, appropriate wedges $\mathcal{W}_\mathcal{D}$ can always be found (see Fig.~\ref{f.causal_wedge}).

Note that this construction is only expected to hold as long as the measurement does not have dramatic effects, that is, the detector couples to an operator in QFT that is at most sub-exponentially complex. This avoids the appearance of event horizons in the bulk as a consequence of the measurement which, in turn, may have drastic repercussions on the existence of the integral kernel required by the causal wedge reconstruction~\cite{Leichenauer13}.

Addressing the second question requires a brief review of information-theoretic tools. In our measurement protocol, we consider a bipartite system consisting of the detector, serving as the apparatus $A$, and the CFT, serving as the system $S$. The mutual information between $S$ and $A$ is 
\begin{equation}
    \mathcal{I}(S:A)=\mathcal{S}[S]+\mathcal{S}[A]-\mathcal{S}[S\cup A]~, 
\end{equation}
with $\mathcal{S}[X]$ denoting the von Neumann entropy of the system $X$~\cite{Zurek83, BarnettP89}. Computing the increase in mutual information shared between the detector and the field due to the coupling interaction is simple in our case, and results\footnote{We use that i) the von Neumann entropy of a pure state is zero, and ii) the von Neumann entropies of two parts of a pure bipartite system always match.} in
\begin{equation}
    \Delta\mathcal{I}(S:A)=2\mathcal{S}[A]~.
    \label{e.delta_I}
\end{equation}
As we will see in a following example, $\mathcal{S}[A]$ is fully determined by the interaction Hamiltonian and CFT data such as the conformal dimension and the OPE coefficients. Moreover, the information-disturbance trade-off tells us that ``\textit{the more useful information is extracted from a quantum system, the more such a system is disturbed by the measurement}''~\cite{BuscemiEtAl08}. This principle is quintessential to quantum information, and has many incarnations across the field~\cite{DAriano03, Maccone06}. While the principle is not easily extended to local measurements in field theory in general~\cite{KretschmannEtAl08}, it is reasonable to expect the disturbance to be lower bounded by the amount of information extracted by the measurement, \textit{i.e.}, by Eq.~\eqref{e.delta_I}. Hence, the information gain tells how much the measurement-induced update affects the boundary's state and, thanks to the holographic duality, how much a boundary measurement affects the bulk. We will return to this point in the conclusions.

Let us show the above construction in a simple example. The interaction Hamiltonian determines the operator applied at the measurement point when the measurement outcome is $m=1$ and, consequently, the post-measurement state of both the boundary and bulk fields. We pick the operator resulting from the system-apparatus coupling to be a primary operator $\hat{\mathcal{O}}_\Delta(x_M)$ of conformal dimension $\Delta$, localised at the measurement point. To yield a normalizable state, we regularise the operator by a small Euclidean time evolution
\begin{equation}
    \hat{\mathcal{O}}_{\Delta,\epsilon}(0,0)=e^{-\epsilon\hat{H}}\hat{\mathcal{O}}_\Delta(0,0)e^{\epsilon\hat{H}}
    \label{e.insertion}
\end{equation}
\cite{BerensteinS20}. This can be seen as adding a small smearing to the boundary interaction, which in turn moves the operator slightly inside the bulk. Because the smearing is small, we will overlook the physical meaning of this operation for the moment and regard it as a regularisation trick only.

Stopping at first order in $\lambda$, measuring the detector and obtaining $m=1$ corresponds to applying the regularised operator at the measurement point. The resulting state is
\begin{equation}
    \ket{\Psi_1; \epsilon}=e^{-\epsilon \hat{H}}\hat{\mathcal{O}}_\Delta(0,0)\ket{0_M}~.
    \label{e.CFT_state_at_boundary}
\end{equation}
We can now study how the information extracted by the boundary measurement encodes the semi-classical properties of the bulk state. To this end, we compute the detector's density operator at the end of the detector-field interaction\footnote{While second order contributions in $\lambda$ matter for obtaining a density operator with proper $\Tr \rho_A =1$ normalization, one can also stop at first order in $\lambda$ and manually fix the state's trace on the $\ket{1}\bra{1}$ component.}:
\begin{equation}
    \rho_A=\begin{pmatrix}
    1-\alpha\lambda^2 &\beta\lambda \\
    \beta^*\lambda  & \alpha\lambda^2 
    \end{pmatrix}~.
    \label{e.perturbed_rho}
\end{equation}
The matrix components in the above expression are the (regularised) primary 's two-point function
\begin{equation}
    \alpha=\bra{0_M}\hat{\mathcal{O}}_\Delta^\dagger(0,0)e^{-2\epsilon\hat{H}}\hat{\mathcal{O}}_\Delta(0,0)\ket{0_M}~,
\end{equation}
and its one-point function $\beta=0$  at the measurement point. Assuming $\alpha\lambda^2\ll 1$~\footnote{This expression for the mutual information is only valid for values of $\lambda^2\alpha$ between zero and $e$. This is satisfied in our analysis, as we assume $\lambda^2\alpha\ll 1$. In practice, this means we are considering the regime where the perturbation parameter is much smaller than the regularisation one. While this condition may fail under a more general analysis, it is always satisfied when the series expansion leading to the density operator \eqref{e.perturbed_rho} is allowed.}, the mutual information reads as
\begin{equation}
    \Delta\mathcal{I}(S:A)\simeq2\alpha\lambda^2\left[1-\log\left(\alpha\lambda^2\right)\right]~.
    \label{e.approx_I}
\end{equation}
and, because $\alpha\simeq(2\epsilon)^{-2\Delta}$, it relates the extracted information to the regularisation parameter and conformal dimension only. Of course, $\Delta\mathcal{I}(S:A)$ would encode even more CFT data at higher orders in perturbation theory.

Moving to the bulk perspective and given that $\Delta\gg 1$ and $\epsilon\ll 1$, it is known that the state \eqref{e.CFT_state_at_boundary} is holographically dual to a bulk field excitation well localised in both the radial and angular directions, with a localisation region of size controlled by the regularisation parameter $\epsilon$~\cite{BerensteinS20}. This can be seen as a semi-classical particle initially at rest at a given location in global AdS. In fact, the mass of the particle is determined by the boundary spacetime dimension ($d=2$ for us) and the conformal dimension of the inserted primary $\Delta$ as
\begin{equation}
    m=\sqrt{\Delta(\Delta-d)}%\simeq\Delta-d/2+O(1/\Delta)~,
    \label{e.Delta_m}
\end{equation}
The particle's classical trajectory in Poincar\'e coordinates %matches the dual CFT state free evolution in Lorentzian time
%\begin{equation}
    %\ket{\Psi_1(t_L)}=e^{-i\hat{H}t_L}\ket{\Psi_1}~,
%\end{equation}
%and 
starts from an insertion point $z_*$ matching the regularization parameter $\epsilon$:
\begin{equation}
    z_*=\epsilon
    \label{e.turning}
\end{equation}
(see App.~\ref{a.AdS_calc}). As the regularised two-point function $\alpha$ is fully determined by $\Delta$ and $\epsilon$ and, in turn, determines the mutual information via \eqref{e.approx_I}, it serves as a bridge to fix the amount of extracted information in terms of the mass and insertion point of the classical particle in AdS via \eqref{e.Delta_m} and \eqref{e.turning}.

\paragraph{An alternative case.} We have discussed an update to a state created by a heavy primary operator, in which case in the bulk a massive particle appears near the boundary and falls in. As a brief alternative example, let the operator coupled to the detector instead be a sum of many light operators\footnote{We thank Samir Mathur for discussions about this point.}. For example, start with the top-down construction with type IIB string theory compactified on $M_{4,1}\times S^1 \times T^4$ with $N_1$ D1-branes wrapped on $S^1$ and $N_5$ D5-branes on $S^1\times T^4$, and consider the low-energy limit where the radius of $S^1$ is large compared to $T^4$. The low-energy limit becomes a 1+1 dimensional CFT living on $S^1$, the D1D5 CFT at the orbifold point, with the symmetrized product target space  $(T^4)^{N_1N_5}/S_{N_1N_2}$. Let the initial state of the CFT be the NS sector vacuum $\ket{0}$ so that the dual theory is IIB supergravity on $AdS_3\times S^3 \times T^4$. Consider a graviton polarized in the directions $X^I,X^J$ of $T^4$ \cite{Lunin:2001jy} (see also {\em e.g.} \cite{Hampton:2019csz}). Upon dimensional reduction to $AdS_3$ the graviton reduces to a massless scalar field together with an infinite tower of massive Kaluza-Klein scalar fields. In the orbifold theory, the operator dual to the bulk field is $O_{IJ} \sim \frac{1}{2}\sum^N_{k=1} \left( \partial_z X^{(k)}_I\partial_{\bar{z}}X^{(k)}_J + (I\leftrightarrow J)\right) $ where $k$ labels the $N=N_1N_5$ copies. In this case, the Schrödinger picture time-evolution after the update would correspond to massless excitations propagating in light-like directions in the CFT, while in the bulk scalar excitations are propagating towards the center. In this case the time-evolution in the CFT would correspond to shockwaves of energy propagating from the measurement point.

\section{Conclusions and Outlook}

We discussed measurements in a relativistic setting and extended the results of \cite{PranziniKV23} to enable a consistent treatment of the standard formulation of QFT in the presence of a measurement. Applying these tools to a conformal field theory with holographic dual, we exploited the AdS/CFT correspondence to translate the measurement-induced state update from the boundary to the bulk, and related information-theoretic quantities at the boundary with semi-classical properties of QFT states in the bulk. The result is that local measurements induce non-local state updates at the boundary, which are then translated to non-local updates in the bulk via the causal wedge reconstruction. This way, the information extracted via detector-based measurements on the boundary corresponds to fixing semi-classical properties of the field state in the bulk. Finally, we provided a brief example, showcasing our construction and results.

Our study has certain limitations. First, we restricted the analysis to point-like measurements involving a sharp detector-field interaction. The implications of using a smeared detector were briefly discussed, and we found that the no-signalling theorem protects against inconsistencies when the detector (and thus its causal past) is smeared. Nonetheless, a more complete treatment should include smearing in both space and time~\cite{Louko06, Satz07, SriramkumarP96}, especially since a preliminary analysis suggest non-trivial bulk behaviour in such cases. Additionally, Ref.~\cite{Leichenauer13} notes that bulk event horizons can significantly affect the existence of the integral kernel required for causal wedge reconstruction, indicating that our framework may not straightforwardly extend to scenarios where measurements strongly alter the bulk causal structure. Of course, the presence of horizons brings additional problems with operators of exponential complexity.

Finally, we treated holography solely as a correspondence between quantum field theories. Specifically, we restricted the analysis to the leading order in $G_N$ for the AdS/CFT correspondence, thereby neglecting any effect of quantum gravity. While this approximation considerably simplifies the analysis and provides a natural starting point, the most interesting phenomena are expected to emerge once quantum gravitational effects are taken into account. 

In the quantum regime of gravity, fluctuations of the causal structure should play a crucial role. Interestingly, hints of such fluctuations already arise when considering smeared detectors, as we briefly mentioned at the end of Sec.~\ref {s.prepare_and_measure_UdW}. However, considering both the fact that (i) the more delocalised the detector at the boundary, the deeper its dual is in the bulk, and (ii) smeared detectors have smeared past light cones, and thus indefinite causal pasts, gives that (iii) delocalised measurements at the boundary may be related to bulk observers far from the boundary, for which the quantum gravitational nature of measurement-related causal structures cannot be overlooked.

Despite these simplifications and the effective restriction of our analysis to a boundary-QFT/bulk-QFT correspondence, the measurement-induced updates presented here can still be used to consider several intriguing scenarios for future work:

{\bf Classical detection of a bulk particle.} Consider the case %presented above 
where a measurement ``injects'' a particle into the bulk of a global AdS$_3$: the particle’s back-reaction cuts off a deficit angle~\cite{Matschull99} which can be detected via classical measurements. % of the deficit angle%In such a setup, 
In the case of a global AdS$_3$ where space is periodic, a boundary observer could infer the particle’s presence by sending a tangential light ray and measuring the shortened round-trip time of the returning ray. 

{\bf Black hole formation by measurements.} Repeated measurements~\cite{PranziniEtAl24} on opposite sides of the boundary cylinder may drive the CFT into a state dual to a black hole. In the bulk, this can (with small probability) lead to a sequence of semi-classical particle insertions, colliding at the centre to form a particle with increasing mass~\cite{Lindgren16}, eventually reaching a regime where black hole formation becomes possible. In this case, the information gained in the measurements should be related to the entropy of the black hole.

{\bf Problems for entanglement wedges.} Recent developments have shown that bulk reconstruction is more efficient when using the entanglement wedge rather than the causal wedge, since boundary data typically encodes information beyond the causal wedge due to the entangled nature of QFT states~\cite{CotlerEtAl19, BoussoP23}. However, using an entanglement wedge in the presence of a measurement may pose problems when the boundary causal diamond lies in the measurement’s past. In such cases, the entanglement wedge may extend beyond the bulk past lightcone emanating from the measurement, reaching into regions where the quantum state has already been updated  (see Fig.~\ref{f.causal_wedge}). This leads to an inconsistency, as the wedge would assign the pre-measurement bulk state to areas where the post-measurement one should instead be used. %We speculate that \sout{this may be due to} local measurements typically disrupting entanglement between the measured subsystem and the rest. 
It would be interesting to understand the analogue of entanglement wedge reconstruction in the presence of measurements.

{\bf Information-disturbance trade-off.} Our preliminary results relating the information extracted by a measurement with the semi-classical properties of the bulk hint at the possibility of describing measurement-induced back-reaction on the gravitational sector in connection with the information gained by measurement. The information-disturbance trade-off says that the more information one extracts through measurements, the greater the disturbance one necessarily induces in the system~\cite{DAriano03}. This perspective naturally leads to speculations about local information retrieval in the presence of an event horizon, and the potential back-reaction measurements may have in this case.

{\bf The experience of a bulk observer.} Suppose an observer in the bulk first sees an empty space, but detects the massive particle in the bulk after a measurement at the boundary. What does the observer experience? Do they gain information and think they had missed detecting a particle, or (as we have understood a comment in the paragraph previous to Eq.~(6.24) in Ref.~\cite{Akers:2022qdl}) is their memory affected to lose information about the past? How would one then quantify the gained/lost information? A natural guess is that it should be equal to the information gained by the measurement at the boundary. It would be interesting to apply this perspective to the newly proposed measurements that can determine the interior state of a black hole \cite{yidirim2025}.

\acknowledgments

The authors thank S. Antonini, S. Mathur, S. Raju, and X. Sun for comments and for drawing our attention to relevant literature. N.P. acknowledges financial support from the Magnus Ehrnrooth Foundation and the Academy of Finland via the Centre of Excellence program (Project No. 336810 and Project No. 336814). N.P. and E.K.-V. acknowledge the financial support of the Research Council of Finland through the Finnish Quantum Flagship project (358878, UH). E.K.-V. is in part supported by the Research Council of Finland grant 1371600.  VB was supported in part by the DOE through DE-SC0013528 and QuantISED grant DE-SC0020360, and in part by the Eastman Professorship at Balliol College, University of Oxford. E.K.-V. gratefully acknowledges support from the Simons Center for Geometry and Physics, Stony Brook University at which some of the research for this paper was performed, during the program ``50 years of black hole information paradox".

% Bibliography

%% [A] Recommended: using JHEP.bst file
 \bibliographystyle{JHEP}
 \bibliography{bib.bib}

\appendix
\section{Slicing of the cylinder}
\label{a.slicing_construction}
In this appendix, we consider a cylindrical spacetime and therein construct a slicing consistent with the presence of a measurement. The spacetime has a flat metric $ds^2=-dt^2+dx^2$, with the identification $x\sim x+2L$. For concreteness, We choose the fundamental domain for the coordinate $x$ to be the interval $x\in(-L,L]$, and we pick the simple case in which the measurement takes place at the origin, \textit{i.e.}, $x_M=(0,0)$. 

To construct the slicing, we start by considering a causal diamond centred at $(t_0,x_0)$ with diagonal $2l$, whose vertices are $x_{\pm,\pm}=(t_0\pm l,x_0\pm l)$. A possible slicing can be obtained by the usual one defined on the compactified Minkowski spacetime as
\begin{equation}
    \Xi_u=\{(t_0+F_+(u;\sigma),x_0+F_-(u;\sigma))~|~\sigma\in[-\pi/2,\pi/2]\}
\end{equation}
with 
\begin{equation}
    F_\pm(u;\sigma)=\frac{l}{\pi}\left\lbrace\arctan[\tan(u) + \tan(\sigma)] \pm \arctan[\tan(u) - \tan(\sigma)]\right\rbrace
\end{equation}
and $u\in[-\pi/2,\pi/2]$ being the time coordinate in the diamond. If instead we want the time parameter to run from $\tau_-$ to $\tau_+$, we can use the change of coordinates
\begin{equation}
    \tau(u)=\frac{\tau_++\tau_-}{2}+\frac{\tau_+-\tau_-}{\pi}u~,
\end{equation}
to get
\begin{equation}
    \Xi_\tau=\{(t_0+\Tilde{F}_+(u(\tau);\sigma),x_0+\Tilde{F}_-(u(\tau);\sigma))~|~\sigma\in(-\pi/2,\pi/2)\}
\end{equation}
with 
\begin{equation}
    \begin{split}
        \mathcal{F}_\pm(\tau;\sigma)=\frac{l}{\pi}\arctan\left[\tan\left(\frac{\tau\pi}{\tau_+-\tau_-}-\frac{\pi}{2}\frac{\tau_++\tau_-}{\tau_+-\tau_-}\right) + \tan(\sigma)\right]
    \\ 
    \pm \frac{l}{\pi}\arctan\left[\tan\left(\frac{\tau\pi}{\tau_+-\tau_-}-\frac{\pi}{2}\frac{\tau_++\tau_-}{\tau_+-\tau_-}\right) - \tan(\sigma)\right]
    \label{a.half_diamond_coordinates_F}
    \end{split}
\end{equation}
and $\tau\in[\tau_-,\tau_+]$; in a similar way, we can trade $\sigma$ for a new variable $\chi$ that runs over any other finite interval $(\chi_-,\chi_+)$:
\begin{equation}
    \begin{split}
        \Tilde{F}_\pm(\tau;\chi)=&\frac{l}{\pi}\arctan\left[\tan\left(\frac{\tau\pi}{\tau_+-\tau_-}-\frac{\pi}{2}\frac{\tau_++\tau_-}{\tau_+-\tau_-}\right) + \tan\left(\frac{\chi\pi}{\chi_+-\chi_-}-\frac{\pi}{2}\frac{\chi_++\chi_-}{\chi_+-\chi_-}\right)\right]\\ \pm &\frac{l}{\pi}\arctan\left[\tan\left(\frac{\tau\pi}{\tau_+-\tau_-}-\frac{\pi}{2}\frac{\tau_++\tau_-}{\tau_+-\tau_-}\right) - \tan\left(\frac{\chi\pi}{\chi_+-\chi_-}-\frac{\pi}{2}\frac{\chi_++\chi_-}{\chi_+-\chi_-}\right)\right]~.
    \end{split}
\end{equation}
%Specifically, in our case we will be interested in diamonds with diagonal $2L$ and in requiring $\chi\in(-L,L)$, the latter needed to make the slices at $\tau=(\tau_++\tau_-)/2$ to have spatial coordinate running from $-L$ to $L$. These requirements give
%\begin{multline}
%    \mathcal{F}_\pm(\tau;\chi)=\frac{L}{\pi}\arctan\left[\tan\left(\frac{\tau\pi}{\tau_+-\tau_-}-\frac{\pi}{2}\frac{\tau_++\tau_-}{\tau_+-\tau_-}\right) + \tan\left(\frac{\pi}{2}\frac{\chi}{L}\right)\right]\\ \pm \frac{L}{\pi}\arctan\left[\tan\left(\frac{\tau\pi}{\tau_+-\tau_-}-\frac{\pi}{2}\frac{\tau_++\tau_-}{\tau_+-\tau_-}\right) - \tan\left(\frac{\pi}{2}\frac{\chi}{L}\right)\right]
%    \label{a.half_diamond_coordinates_F_final}
%\end{multline}
%and
%\begin{equation}
%    \Sigma(\tau)=\{(t_0+\mathcal{F}_+(\tau;\chi),x_0+\mathcal{F}_-(\tau;\chi))~|~\chi\in(-L,L)\}~.
%\end{equation}
When placed on the unwrapped cylinder, a causal diamond can overlap itself. This happens if its diagonal $2l$ is longer than the cylinder circumference $2L$. In such a case, the slicing must be adjusted to smoothly run across the identified boundary. Specifically, matching the Minkowski slices at the crossing point will create symmetrical cusps, which require smoothing.

We are now ready to construct our slicing. The first past lightcone intersection are at $x_I=(-L,-L)\sim(-L,L)$. We respectively call $t_M$ and $t_I$ the coordinate times of the measurement and of the past lightcone intersections, \textit{i.e.}, $t_M=0$ and $t_I=-L$. Next, pick two positive numbers $\delta_+$ and $\delta_-$ and we divide our spacetime in the following regions: i) the region with coordinate time $t<t_I-\delta_-$; ii) the region with coordinate time $t>t_M+\delta_+$, for which the usual slicing can be used (blue regions in the figure below); iii) a region containing all remaining points in the measurement's past lightcone (red region below); and iv) a region containing all remaining points outside of the measurement's past for which the usual slicing is not available (yellow region below). While $\delta_\pm$ can be sent to zero in the above definitions, it is more useful to initially keep them finite, and later check if the limit can be performed in a meaningful way.

\begin{figure}[]
  \centering
  % Subfigure with TikZ drawing
  \begin{subfigure}[b]{0.45\textwidth}
    \centering
\hspace*{-1.25cm}

\begin{tikzpicture}[scale=1.4, line cap=round, line join=round]

% Define pastel colors
\definecolor{pastelcyan}{RGB}{178, 235, 242}  % Soft Cyan
\definecolor{pastelyellow}{RGB}{255, 249, 196} % Soft Yellow
\definecolor{pastelpink}{RGB}{255, 204, 204}   % Soft Red/Pink

% Coordinates for the triangle vertices
\coordinate (Xt) at (0,2);   % Top vertex
\coordinate (Xl) at (-2,0);  % Left vertex
\coordinate (Xr) at ( 2,0);  % Right vertex

\coordinate (A) at ( -2,2);
\coordinate (B) at ( 2,2);

\coordinate (Ap) at ( -2,2.4);
\coordinate (Bp) at ( 2,2.4);
\coordinate (Xlp) at (-2,-0.4);  % Left vertex
\coordinate (Xrp) at ( 2,-0.4);  % Right vertex

% Fill areas with pastel colors
\fill[pastelcyan] (-2,2) rectangle (2,3);  % Top cyan rectangle
\fill[pastelcyan] (-2,0) rectangle (2,-1); % Bottom cyan rectangle
\fill[pastelyellow] (A) -- (Xl) -- (Xt) -- cycle;  % Left triangle (yellow)
\fill[pastelyellow] (B) -- (Xr) -- (Xt) -- cycle;  % Right triangle (yellow)
\fill[pastelyellow] (-2,2) rectangle (2,2.4);
\fill[pastelpink] (Xl) -- (Xr) -- (Xt) -- cycle;   % Center triangle (pink/red)
\fill[pastelpink] (Xlp) rectangle (Xr);

% Dashed vertical lines
\draw[dash dot] (-2,-1) -- (-2,3);
\draw[dash dot] ( 2,-1) -- ( 2,3);

% Draw the triangle edges in black
\draw[thick, black] (Xt) -- (Xl);
\draw[thick, black] (Xt) -- (Xr);
\draw[dashed, black] (Xl) -- (Xr);
\draw[thick, black] (Xlp) -- (Xrp);
\draw[dashed, black] (A) -- (B);
\draw[thick, black] (Ap) -- (Bp);

% Example of labeling the triangle vertices (optional)
\node at (Xt) [above] {$x_M$};
\node at (Xl) [below left] {$x_I$};
\node at (2,-0.07) [below right] {$\delta_-$};
\node at (2,2.35) [below right] {$\delta_+$};

% Draw black points at Xl, Xr, Xt
\fill[black] (Xt) circle (2pt);
\fill[black] (Xl) circle (2pt);
\fill[black] (Xr) circle (2pt);

\end{tikzpicture}
\label{f.patching}
    \caption{Regions which require different slicings.}
  \end{subfigure}
  \hfill
  % Subfigure with PNG image
  \begin{subfigure}[b]{0.45\textwidth}
    \centering
    \includegraphics[width=\textwidth]{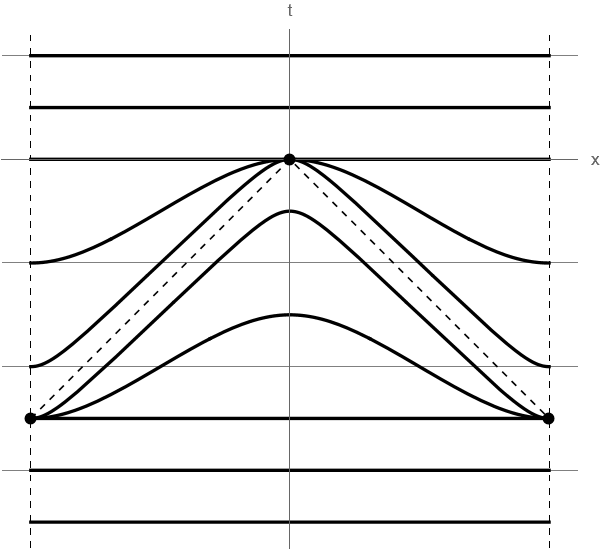
    }
    \caption{Slicing of the cylinder ($\delta_+=\delta_-=0$).}
  \end{subfigure}
  \caption{Construction of a slicing respecting the spacetime regions in which a field state can be properly assigned when a measurement is performed at $x_M$.}
  \label{fig:subfigure}
\end{figure}
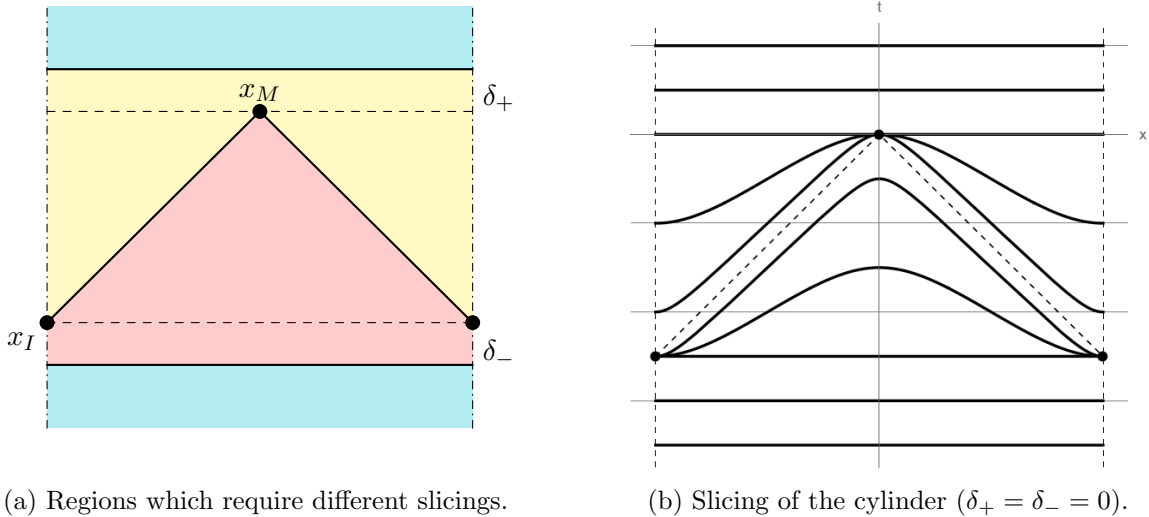

Then we introduce the slices
\begin{equation}
    \Sigma(\tau)=\{(\tau,x)~|~x\in(-L,L]\}~;
\end{equation}
with $\tau\leq t_I-\delta_-$ and $\tau\geq t_M+\delta_+$, which cover the first two regions above. Then we paste two half-diamonds, $D_r$ and $D_y$, onto the spacetime, and use their coordinates to describe the covered parts of the cylinder. We take the former to have upper vertex at $x_M$ and diagonal $d_r=|t_I|+\delta_-$, \textit{i.e.} it fully covers the red region above. Similarly, we take the latter with lower vertex at $x_I$ and diagonal $d_y=|t_I|+\delta_+$, \textit{i.e.} it fully covers the yellow region above. The slices of these half-diamonds should cover the values of $\tau$ from $\tau=t_I-\delta_-=-(L+\delta_-)$ to $\tau= t_M+\delta_+=\delta_+$. For the sake of simplicity, we take $\delta_+=\delta_-=\delta$, and split the available $\tau$ range to uniformly cover the two regions. Thus, we let $\tau\in(-(L+\delta), -L/2)$ to cover the red region, and $\tau\in(-L/2,\delta)$ to cover the yellow region; notice that both regions have span a ``duration'' $L/2+\delta$ in $\tau$-time. Next, we send $\delta\to 0$ to avoid the causal diamonds having diagonal longer than $2L$. The resulting half-diamonds are such that: 
\begin{enumerate}
    \item the red one, called $D_r$, is half of a diamond centred at $(-L, 0)$, has sides of length $2\sqrt{2}$, and its time runs from $-L$ on the bottom space like line to $-L/2$ on the top light like surface;
    \item the yellow one, called $D_y$, is half of a diamond centred at $(0, L)$, has sides of length $2\sqrt{2}$, and its time runs from $-L/2$ on the bottom light like surface to $0$ on the top space like line.
\end{enumerate} 
Having these information, we can use the above general discussion to find the specific details of our slicing. Specifically,
\begin{equation}
    F^p_\pm(\tau;\chi)=\frac{L}{\pi}\left\lbrace\arctan\left[\tan\left(\frac{\pi\tau}{L}\right)+\tan\left(\frac{\pi\chi}{2L}\right)\right] \pm \arctan\left[\tan\left(\frac{\pi\tau}{L}\right)-\tan\left(\frac{\pi\chi}{2L}\right)\right]\right\rbrace
\end{equation}
and 
\begin{equation}
    F^f_\pm(\tau;\chi)=\frac{L}{\pi}\left\lbrace\arctan\left[\tan\left(\frac{\pi\tau}{L}\right)+\cot\left(\frac{\pi\chi}{2L}\right)\right] \pm \arctan\left[\tan\left(\frac{\pi\tau}{L}\right)-\cot\left(\frac{\pi\chi}{2L}\right)\right]\right\rbrace~.
\end{equation}
Finally, the slicing results in
\begin{equation}
    \Sigma(\tau)=\begin{dcases}
        \{(\tau,\sigma)~|~\sigma\in(-L,L]\}~~&{\rm for}~~\tau\leq-L\\
        \{(-L+F^p_+(\tau,\sigma),F^p_-(\tau,\sigma))~|~\sigma\in(-L,L]\}~~&{\rm for}~~-L< \tau<-L/2\\
        \{(F^f_+(\tau,\sigma),{\rm sign}(\sigma)L+F^f_-(\tau,-\sigma))~|~\sigma\in(-L,L]\}~~&{\rm for}~~-L/2<\tau<0\\
        \{(\tau,\sigma)~|~\sigma\in(-L,L]\}~~&{\rm for}~~\tau\geq0
    \end{dcases}
\end{equation}
with the additional time like slice $\Sigma(-L/2)=\{(-|\sigma|,\sigma)~|~\sigma\in(-L,L]\}$. Notice that, in the above formula, the sign function ${\rm sign}(\sigma)$ is required for $\sigma$ to correctly span from $-L$ to $L$ inside the range $(-L,L)$ in $(t,x)$ coordinates, and the minus sign inside $F^f_-$ is required for it to not wrap around the cylinder at $\sigma=0$.

%In the mapping \eqref{e.Lor_to_Euc_map} from points on the Lorentzian plane to those on the radial Euclidean plane in the coordinates $(z,\Bar{z})$, the slicing reads
%\begin{equation}
%    \Sigma(\tau)=\begin{dcases}
%        \{\exp( \tau+i \sigma\pi/L)~|~\sigma\in(-L,L]\}~~&{\rm for}~~\tau\leq-L\\
%        \{\exp(-L+F^p_+(\tau,\sigma)+i\pi F^p_-(\tau,\sigma)/L)~|~\sigma\in\mathbb{R}\}~~&{\rm for}~~-L< \tau\leq-L/2\\
%        \{\exp(F^f_+(\tau,\sigma)+i\pi({\rm sign}(\sigma)L+F^f_-(\tau,-\sigma))/L)~|~\sigma\in\mathbb{R}\}~~&{\rm for}~~-L/2<\tau<0\\
%        \{\exp( \tau+i \sigma\pi/L)~|~\sigma\in(-L,L]\}~~&{\rm for}~~\tau\geq0
%    \end{dcases}
%    \label{e.slices_E}
%\end{equation}

\section{Local vs. global perspectives of measurements in CFT}
\label{a.CFT_details}

As described in the main text, in Minkowski spacetime, we obtain the $\tau$-slicing by constructing coordinate charts spanning $\mathcal{R}_P$ and $\mathcal{R}_I$ as follows. First, we use past Rindler coordinates $(\eta,\xi)$ to cover $\mathcal{R}_P$. These are related to $(t,x)$ via
\begin{equation}
    \begin{dcases}
        t=-e^{\eta}\cosh\xi\\
        x=-e^{\eta}\sinh\xi
    \end{dcases}
    \label{e.Milne_coordinates}
\end{equation}
with $\eta,\xi\in\mathbb{R}$. This shows that the past of the measurement can be seen as a contracting Milne spacetime. Next, we cover the left and right portions of the region $\mathcal{R}_I$ with left/right Rindler coordinates 
\begin{equation}
    \begin{dcases}
        t=\rho\sinh\xi\\
        x=(-)\rho\cosh\xi
    \end{dcases}
\end{equation}
with $\rho>0$ and $\xi\in\mathbb{R}$, and where the minus sign in parentheses applies to the left region only. Since our discussion will primarily focus on $\mathcal{R}_P$, we will not consider the $\mathcal{R}_I$ region further. Nevertheless, it is worth noting that quantisation in $\mathcal{R}_I$ can be carried out by standard techniques of QFT in non-inertial frames~\cite{Unruh76}.

Focusing on $\mathcal{R}_P$, the coordinate system makes it evident that it is a flat contracting universe, on which QFT can be solved analytically~\cite{BirrellD82}. The coordinates \eqref{e.Milne_coordinates} give the metric
\begin{equation}
    ds^2=e^{2\eta}(-d\eta^2+d\xi^2)~,
    \label{e.Milne_metric}
\end{equation}
which is conformally flat, and has two conformal time-like future-directed Killing vector fields 
\begin{equation}
    \begin{dcases}
        K_1=-\partial_\eta\\
        K_2=-e^{-\eta}(\cosh\xi\partial_\eta-\sinh\xi\partial_\xi)
    \end{dcases}
\end{equation}
which define two distinct vacua $\ket{0_1}$ and $\ket{0_2}$, respectively called the conformal and adiabatic vacuum~\cite{CandelasD79, VaudrevangeK07}. Also, the vector fields define two distinct collections of spacelike hypersurfaces. The first is the collection of surfaces with fixed $\eta$, and the second is the restriction to $\mathcal{R}_P$ of the usual Minkowski inertial slices (for $K_2=\partial_t$). Moreover, the two slicings differ in the closure of the evolution they generate. The first slicing can be viewed as a conformal rescaling to the origin's past of the usual inertial slicing, and thus generates a closed evolution within $\mathcal{R}_P$: imposing a pure state on some initial slice means finding a pure state on later slices also. In particular, $\ket{0_1}$ stays invariant along the evolution generated by $K_1$ when imposed on an initial slice. In contrast, $K_2$ does not preserve the region $\mathcal{R}_P$ and therefore no state that is (globally) invariant under $K_2$ can remain pure when restricted to $\mathcal{R}_P$. Indeed, $\ket{0_2}=\ket{0_M}$ is globally invariant along the evolution generated by $K_2$ and pure on the full Minkowski spacetime, but its restriction to $\mathcal{R}_P$ is a mixed state---specifically, the thermal state with respect to the time evolution generated by $K_1$~\cite{OlsonR11}.

%\subsection{Is the local perspective compatible with the global one?}
%\label{a.local_vs_global}
Let us now understand how the usual measurement picture can be reconciled with the experience of a local observer. To this end, we must select an initial state, explicitly examine time evolution along the slices within $\mathcal{R}_P$, and track how the state---defined at $\tau=-\infty$, where $\Sigma(\tau)$ coincides with a portion of the usual $t_l=const.$ slicing---evolves towards the measurement. 
Because the past wedge is described by the metric \eqref{e.Milne_metric}, quantisation of a massless field in $\mathcal{J}^-(x_M)$ is equivalent to that in the contracting Milne spacetime (or, equivalently, the past Rindler wedge). Focusing on the massless case, we exploit the fact that the time evolution generated by $K_1$ preserves the restriction to $\mathcal{R}_P$ of the vacuum state defined by $K_2$ (i.e., a partial trace of the Minkowski vacuum), and thus the state seen by the measuring observer on the final slice before the measurement still is (a restriction of) the inertial vacuum, thus making the above measurement description valid. 

To make this concrete, we begin with the description of a scalar field in Minkowski via light-cone coordinates. Defining $U=t-x$ and $V=t+x$, we denote by $\ket{0_M}$ the vacuum of the $t$-positive frequency left- and right-moving modes, \textit{i.e.},
\begin{equation}
    \hat{a}^{L}_k\ket{0_M}=\hat{a}^{R}_k\ket{0_M}=0~,
\end{equation}
where the annihilation operators $\hat{a}^{L/R}_k$ are defined by
\begin{equation}
    \hat{\phi}(U,V)=\int_0^\infty \frac{dk}{\sqrt{4\pi  k}}\left(\hat{a}^L_ke^{-ik U}+\hat{a}^R_ke^{-ik V}+h.c.\right)~.
\end{equation}
Next, we consider the light-cone coordinates in the causal past of the origin constructed from the variables \eqref{e.Milne_coordinates} as $u=-\eta+\xi$ and $v=-\eta-\xi$, which give the metric
\begin{equation}
    ds^2=-e^{-v-u}dudv
\end{equation}
and the equations of motion
\begin{equation}
    \partial_u\partial_v\phi=0~.
\end{equation}
These give a new vacuum annihilated by the $\eta$-positive frequency left- and right-moving modes, \textit{i.e.},
\begin{equation}
    \hat{p}^{L}_k\ket{0_1}=\hat{p}^{R}_k\ket{0_1}=0~,
\end{equation}
where the annihilation operators $\hat{p}^{L/R}_k$ are defined by
\begin{equation}
    \hat{\phi}(u,v)=\int_0^\infty \frac{dk}{\sqrt{4\pi  k}}\left(\hat{p}^L_{k}e^{-ik u}+\hat{p}^R_{k}e^{-ik v}+h.c.\right)~.
\end{equation}
As already implicit in the notation we used, the modes $e^{-iku}$ and $e^{-ikv}$ are positive-frequency with respect to $K_1$ and thus define the conformal vacuum $\ket{0_1}$. As it was shown in \cite{OlsonR11}, these modes can be extended to the left/right Rindler wedges, but, to achieve completeness, one must also supplement them with a set of positive frequency modes supported in the future of the measurement (which also extend to the left/right Rindler wedges, and are related to the creation/annihilation operators $\hat{f}^{L/R}_k$ and $\hat{f}^{L/R\dagger}_k$). Still, the time evolution generated by $K_1$ in $\mathcal{R}_P$ is given in terms of past-supported operators only as
\begin{equation}
    \hat{H}_P=\int_0^\infty dk \,k(\hat{p}^{L\dagger}_{k}\hat{p}^L_{k}+\hat{p}^{R\dagger}_{k}\hat{p}^R_{k})~,
\end{equation}
where normal ordering is understood.

An observer who wants to measure by the procedure outlined in Sec.~\ref{s.prepare_and_measure_UdW} will start by introducing a quantisation scheme in $\mathcal{R}_P$ and imposing some state at early times. Because the vacua defined by $K_1$ and $K_2=\partial_t$ are distinct, an early local preparation of the former is not equivalent to an early global preparation of the latter. In fact, the Minkowski vacuum $\ket{0_M}$ is mixed from the perspective of the quantisation scheme defined by $K_1$. Yet, the resulting state is invariant under the time evolution $\hat{U}_P(\tau)=\exp(-i\hat{H}_P \tau)$ generated by $K_1$:
\begin{equation}
    \rho_{\mathcal{R}_P}\mapsto \hat{U}_P(\tau)\rho_{\mathcal{R}_P}\hat{U}^\dagger_P(\tau)=\rho_{\mathcal{R}_P}~.
\end{equation}
Thus, one can use the quantisation defined by $K_1$ and the vacuum defined by $K_2$ and still get a trivial time evolution. Moreover, the final state $\rho_{\mathcal{R}_P}$, supported infinitiesimally near $\gamma$ when approaching $\tau=\tau_*$ from below, can be extended to the initial slice of $\mathcal{R}_I$ and evolved along the slicing therein, giving $\ket{0_M}$ at $\tau=0$.

Given this, we can focus our attention on the measurement process seen as a transition from the initial state $\ket{0_M}$ to the state on the slice $\Sigma_0$, described by an insertion at $x_M$ of the measurement operator conditioned on the measurement outcome---\textit{i.e.},
\begin{equation}
    \ket{0_M}\rightarrow\hat{O}_\Phi(x_M)^m\ket{0_M}=\ket{\Psi_m}
    \label{e.insertion_def_in_APP}
\end{equation}
with $m=0,1$---where we used that the state must be rescaled and cleared of any overall phase factors after post-selection. 

\section{Localised excitations from measurements on the plane}
\label{a.AdS_calc}

Let us consider the boundary operator $\hat{O}_\Delta$ regularised by small euclidean time evolution: $\hat{O}_{\Delta,\epsilon}=\hat{O}_{\Delta}(i\epsilon,0)$. Using the Poincar\'e patch (with AdS radius set to unity), in the bulk this reads as
\begin{equation}
    \hat{O}^{{\rm Bulk}}_{\Delta,\epsilon}(z,Y)=\int d^2x K(z,Y|x)\hat{O}_{\Delta,\epsilon}(x)
\end{equation}
where $K(z,Y|x)$ is the  bulk-to-boundary propagator for a scalar field of conformal dimension $\Delta$, \textit{i.e.}
\begin{equation}
    K(z,Y|x)=\left(\frac{z}{z^2+|Y-x|^2}\right)^\Delta
\end{equation}
Putting everything together, we obtain
\begin{equation}
    \hat{O}^{{\rm Bulk}}_{\Delta,\epsilon}(z,Y)=\left(\frac{z}{z^2+|Y|^2+\epsilon^2}\right)^\Delta\hat{O}_{\Delta,\epsilon}~,
\end{equation}
where we used $\hat{O}_{\Delta,\epsilon}(t,x)=\delta(x)\delta(t-i\epsilon)\hat{O}_{\Delta,\epsilon}$. Evaluating at $Y_0=Y_1=0$ gives the radial profile of the bulk operator
\begin{equation}
    \hat{O}^{{\rm Bulk}}_{\Delta,\epsilon}(z,0)\sim\left(\frac{z}{z^2+\epsilon^2}\right)^\Delta~,
\end{equation}
which peaks at $z_*=\epsilon$, that is, near the boundary of AdS. For small $\epsilon$ and large $\Delta$, the operator's profile is also strongly peaked around its maximum, thus allowing an interpretation of bulk field excitation generated by the boundary measurement as a semi-classical particle localised near the boundary. For this reason, let us now consider the action of a point particle of mass $m$ in AdS$_3$ in Poincar\'e coordinates
\begin{equation}
    S=-m\int d\lambda\sqrt{-g_{\mu\nu}\frac{dx^\mu}{d\lambda}\frac{dx^\nu}{d\lambda}}=-m\int \frac{dt}{z(t)}\sqrt{1-\Dot{z}^2(t)-\Dot{x}^2(t)}
\end{equation}
(in the $\lambda=t(\lambda)$ gauge) and search for the energy for a radial trajectory with turning point $z_*=\epsilon$. From the Lagrangian
\begin{equation}
    L(z,\Dot{z})=-\frac{m}{z(t)}\sqrt{1-\Dot{z}^2(t)}~,
\end{equation}
we obtain the radial momentum
\begin{equation}
    \pi_z(t)=\frac{m}{z}\frac{\Dot{z}}{\sqrt{1-\Dot{z}^2}}
\end{equation}
and the Hamiltonian
\begin{equation}
    H=\frac{m}{z}\left(\frac{\Dot{z}}{\sqrt{1-\Dot{z}^2}}+\sqrt{1-\Dot{z}^2}\right)~.
\end{equation}
Evaluating at the turning point $z(t_*)=z_*=\epsilon$ and $\Dot{z}(t_*)=0$ gives the energy of the point particle:
\begin{equation}
    E=\frac{m}{\epsilon}~.
\end{equation}

%% or
%% [B] Manual formatting (see below)
%% (i) We suggest to always provide author, title and journal data or doi:
%% in short all the informations that clearly identify a document.
%% (ii) please avoid comments such as "For a review'', "For some examples",
%% "and references therein" or move them in the text. In general, please leave only references in the bibliography and move all
%% accessory text in footnotes.
%% (iii) Also, please have only one work for each \bibitem.

\end{document}